\documentclass[aps, 
amsmath, showpacs, preprintnumbers, twocolumn, amssymb, amsmath, superscriptaddress]{revtex4-1}
\usepackage{graphicx}
\usepackage{epstopdf}
\usepackage{mathptmx, textcomp}
\usepackage[latin1]{inputenc}
\usepackage{tabularx}
\usepackage{units}
\usepackage{color}
\usepackage{bm}

\begin{document}
\title{Spontaneous continuous orbital motion of a pair of nanoparticles levitated in air}
\author{Mitsuyoshi\,Yoneda}
\affiliation{Department of Physics, Tokyo Institute of Technology, Ookayama 2-12-1, Meguro-ku, 152-8550 Tokyo}
\author{Makoto\,Iwasaki}
\affiliation{Department of Physics, Tokyo Institute of Technology, Ookayama 2-12-1, Meguro-ku, 152-8550 Tokyo}
\author{Kiyotaka\,Aikawa}
\affiliation{Department of Physics, Tokyo Institute of Technology, Ookayama 2-12-1, Meguro-ku, 152-8550 Tokyo}

\date{\today}

\pacs{}

\begin{abstract}
We report on the discovery of a unidirectional continuous orbital motion of a single pair of nanoparticles which occurs spontaneously in room-temperature air and can be manipulated by light. By varying the relative position of two nanoparticles, we demonstrate a phase transition between two Brownian particles and a pair of co-orbiting particles. The orbital motion is sensitive to air pressure and is vanishing at low pressure, suggesting that the orbital motion is supported by air. Our results pave the way for manipulating nanoscale objects on the basis of their cooperative dynamics.
\end{abstract}

\maketitle

Nanoscale mechanical devices play crucial roles in the operation of biological systems~\cite{schliwa2006molecular,kinbara2005toward}. Recent years have witnessed substantial progresses in designing and manufacturing artificial versions of such devices, resulting in a plethora of molecular motors and rotors~\cite{kinbara2005toward,kay2007synthetic}. These devices comprise functional molecular parts, with each part subject to random Brownian motion imposed by surrounding environment. Extensive efforts in biophysics and nonequilibrium physics have elucidated that the Brownian motion triggers unidirectional transport of nanoscale objects~\cite{astumian1998fluctuation,reimann2002brownian,hanggi2009artificial,faucheux1995optical}. 

A distinct approach for manipulating nanoscale objects has been established in the field of optomechanics, where the motion of objects is precisely controlled via light-matter interactions~\cite{aspelmeyer2014cavity,kippenberg2008cavity}. With the aim of engineering mesoscopic and macroscopic quantum states, a number of studies have realized the coherent control of the motion of nano- and micro-mechanical oscillators, such as cooling to their quantum ground state~\cite{o2010quantum,teufel2011sideband,chan2011laser}. In these studies, in stark contrast to biological and artificial nanomachines, dissipations to the surrounding environment are major obstacles in controlling the objects and thus are carefully removed, e.g. by evacuating the chamber and/or by pre-cooling the entire system. 

\begin{figure}[t]
\includegraphics[width=0.95\columnwidth] {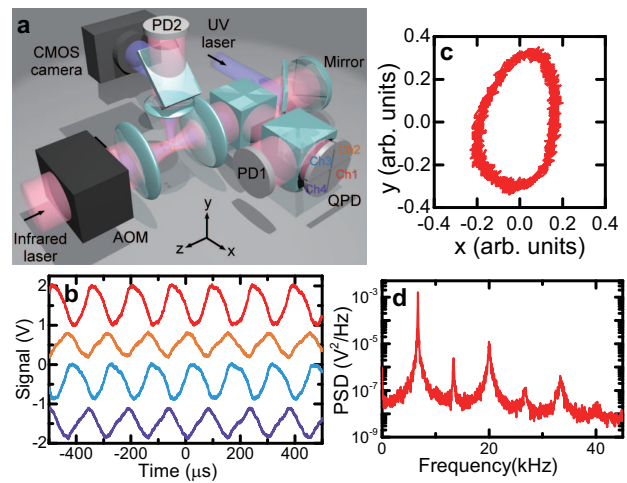}
\caption{(color online). Observation of spontaneous continuous oscillations with nanoparticles trapped in air. (a) Schematic illustration of our experimental setup~\cite{fnote5}. Nanoparticles are trapped in an optical lattice formed with a single-frequency infrared laser. By detecting the scattered infrared light, we measure the motion of nanoparticles. Trapped nanoparticles are imaged on a CMOS camera with an ultraviolet (UV) light. (b) Oscillation signals from the QPD. The signals from the channel 1, 2, 3, and 4 are vertically shifted and aligned from top to bottom. (c) Projection of the trajectory of nanoparticles on the $xy$ plane recovered from (b). (d) PSD calculated from the time trace of PD1 acquired simultaneously with (b).  }
\label{fig:1}
\end{figure}

Here, we report on the discovery and the control of spontaneous continuous optomechanical oscillations of nanoparticles laser-trapped in room-temperature air. By measuring the spatiotemporal evolution of the light scattered by nanoparticles, we clarify that the oscillation originates from a continuous orbital motion of nanoparticles. Furthermore, we reveal that the orbital motion occurs only when more than one nanoparticles are trapped in a single optical potential and stops when they are spatially separated. The observed phase transition between two Brownian particles and a pair of co-orbiting particles is a particular realization of the collective pattern formation of stochastically moving objects, which have been of great interest in biological, physical, and social systems~\cite{romanczuk2012active}. A remarkable aspect of the orbital motion is that it is relevant to both of the aforementioned two approaches in controlling nanoscale objects: the orbital motion occurs only in a strongly dissipative environment, while the orbiting frequency is proportional to the laser power and can be precisely controlled by it. The observed process is qualitatively different from known light-induced oscillations such as phonon lasing~\cite{vahala2009phonon,grudinin2010phonon} and the parametric instability~\cite{braginsky2001parametric,rokhsari2005radiation,metzger2008self}, which exhibit a threshold and a saturation behavior with respect to the power of the applied light and does not require any dissipative mechanism. 

Our experimental setup is based on experiments with levitated nanoparticles~\cite{li2011millikelvin,gieseler2012subkelvin,vovrosh2017parametric,yoneda2017thermal} and is schematically shown in Fig.\,\ref{fig:1}(a)~\cite{fnote5}. We trap nanoparticles in a standing-wave optical trap, an optical lattice, that has been commonly used in atomic physics experiments~\cite{bloch2008many}. Because of the deep optical potential of the order of $k_{\rm B}\times\unit[1\times10^4]{K}$, where $k_{\rm B}$ is the Boltzmann constant, trapped nanoparticles stay in a single site of the optical lattice for hours. Specifically for the present study, we introduce two components. First, a high-numerical-aperture objective lens allows us to acquire site-resolved images of trapped nanoparticles. Second, a quadrant photodetector (QPD) allows us to measure the spatiotemporal evolution of the motion of nanoparticles~\cite{millen2014nanoscale}.

\begin{figure}[t]
\includegraphics[width=0.95\columnwidth] {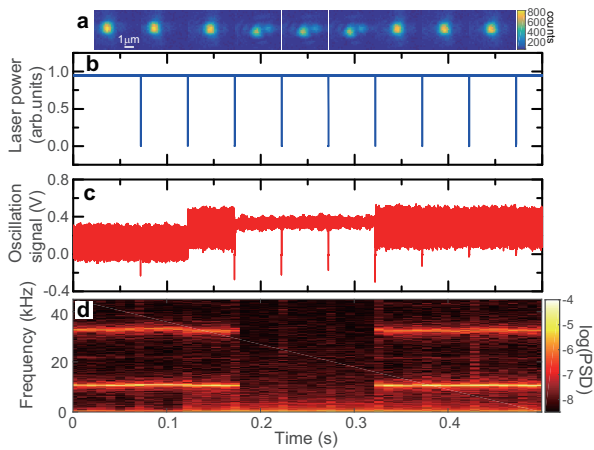}
\caption{(color online). Phase transition between two Brownian particles and a pair of co-orbiting nanoparticles.  (a) Series of the images of nanoparticles trapped in an optical lattice when the trapping beam is switched off for $\unit[400]{\mu s}$ every $\unit[50]{ms}$. The optical lattice has a spacing of $\unit[775]{nm}$. (b) Time variation of the power of the trapping laser. Each image in (a) is acquired $\unit[25]{ms}$ before each switching-off pulse with an exposure time of $\unit[5]{ms}$. (c) Time trace of the oscillation signal. (d) Time variation of the PSDs calculated from (c). }
\label{fig:2}
\end{figure}

The motion of nanoparticles trapped in air is random due to their Brownian motion. It is only at low pressures of below $\unit[100]{Pa}$ that their motion shows a clear oscillation with the frequency of the harmonic confinement~\cite{li2011millikelvin,gieseler2012subkelvin,vovrosh2017parametric,yoneda2017thermal,grimm2000optical,chang2010cavity}. Previous experiments for cooling the motion of nanoparticles have been performed in such a pressure regime and have employed silica nanoparticles with radii of about $\unit[70]{nm}$. By contrast, the present study uses ${\rm Cu}_2{\rm O}$ nanoparticles with radii of about $\unit[80]{nm}$. We discover that, when trapped in an optical lattice, they spontaneously exhibit a clear oscillatory behavior despite in room-temperature air~\cite{fnote4}. Once the oscillation starts, it survives for more than two hours without changing its frequency and amplitude. The phase difference of about 90 degrees among four channels of the QPD unambiguously shows that trapped nanoparticles are orbiting [Fig.\,\ref{fig:1}(b)]. The trajectory of the orbiting nanoparticles, recovered from the QPD signals, is shown in Fig.\,\ref{fig:1}(c)~\cite{fnote5}.  The phase relations are stable for hours, indicating that the orbital motion is unidirectional and its axis is fixed~\cite{fnote3}. The orbiting frequencies observed during many realizations are scattered widely from a few kHz up to $\unit[400]{kHz}$ and do not coincide with the frequencies of the harmonic confinement~\cite{fnote5}. The temperature of the orbital motion calculated from its amplitude is of the order of $\unit[1000]{K}$, suggesting that there exists a robust mechanism for supporting the far-from-equilibrium dynamics. 

It is very unlikely that single Brownian particles abruptly start to orbit at a specific frequency and keep orbiting stably. With a speculation that more than one nanoparticle is involved in this phenomenon, we test if it is possible to release one of trapped nanoparticles by applying an intensity modulation on the trapping beam. As a result, we find that switching off the trapping beam induces the locomotion of trapped nanoparticles along the trapping beam and occasionally alters the relative distance of two nanoparticles (Fig.\,\ref{fig:2}). After the trapping beam is switched off for $\unit[400]{\mu s}$, two nanoparticles trapped in a single lattice site get apart and trapped separately in two lattice sites. At the same instance, the orbital motion is stopped. In this configuration, the wide-spread power spectrum indicates that there are only Brownian particles. With additional switching-off pulses, they again get trapped in a single lattice site and start to orbit at the same frequency as before. Such a behavior is the direct evidence for the two crucial aspects of the phenomenon. First, the orbital motion occurs only when more than one nanoparticles are trapped in a single site of the optical lattice. Second, even trapped in the same potential, these nanoparticles are not attached to each other and can move individually~\cite{fnote1}.
 
\begin{figure}[t]
\includegraphics[width=0.95\columnwidth] {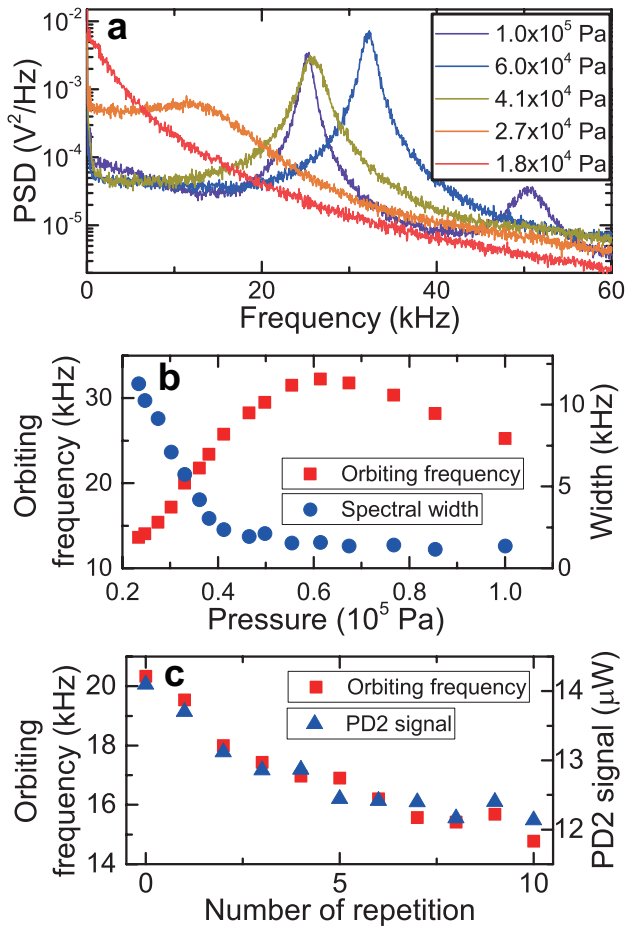}
\caption{(color online). Variation of the orbiting frequency with respect to the pressure. (a) PSDs of the orbiting nanoparticles at various pressures. (b) Orbiting frequencies (left axis) and spectral widths (right axis) with respect to the pressure. (c) Orbiting frequencies after multiple repetitions of decreasing the pressure to around $1\times \unit[10^4]{Pa}$ and increasing to $4.9\times \unit[10^4]{Pa}$. The power of the light scattered by nanoparticles is also shown in the right axis. }
\label{fig:3}
\end{figure}

The striking aspect of the orbital motion is that it survives without changing its amplitude for hours even in the presence of air friction. For clarifying the role of air, we vary the pressure around the trap region and acquire the oscillation signals at various pressures [Fig.\,\ref{fig:3}, (a) and (b)]. We extract the orbiting frequencies and the spectral widths by fitting the PSD with a Lorentzian function~\cite{fnote5}. Remarkably, we observe a behavior opposite to our expectation of observing even stronger oscillations at low pressures. The orbiting frequency shows a maximum value at around $6\times \unit[10^4]{Pa}$. Below this pressure, the orbital motion is slowed with a decreased pressure and is hardly observable at below $2\times \unit[10^4]{Pa}$. Approaching this pressure, we observe a dramatic increase in the spectral width of the PSD. These facts show that the presence of air is essential for the orbital motion.

  It is interesting to see what would happen to the nanoparticles if we bring them to a low pressure, where the orbital motion is hardly observable, and then again to higher pressures. In particular, we are interested in whether the orbiting frequency is reproduced each time. The result of repeating such a procedure is shown in Fig.\,\ref{fig:3}(c), where the measured orbiting frequency at a specific pressure ($4.9\times \unit[10^4]{Pa}$) is plotted with respect to the number of repetitions. Importantly, we observe a slight, but clear monotonic decrease in the orbiting frequencies with each repetition. We interpret that this decrease results from a decrease in the sizes of nanoparticles. As shown in Fig.\,\ref{fig:3}(c), the scattered infrared light also decreases with repetition. The magnitude of Rayleigh scattering from a nanoscale object is proportional to $a^6$ with $a$ being the radius of nanoparticles~\cite{cox2002experiment} and thus is a sensitive measure of the size of nanoparticles. We infer that bringing nanoparticles to low pressures promotes the desorption of gas molecules attached to nanoparticles and accordingly alters the orbiting frequency. This measurement indicates that the orbiting frequency strongly depends on the size of nanoparticles.

\begin{figure}[t]
\includegraphics[width=0.95\columnwidth] {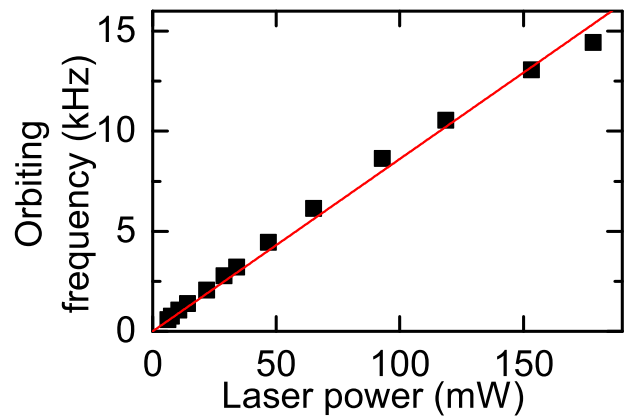}
\caption{(color online). Variation of the orbiting frequency with respect to the laser power at atmospheric pressure. The solid line is a linear fit on the data points without intercept. }
\label{fig:4}
\end{figure}

An important insight on the orbiting process is obtained by measuring how it is influenced by the trapping laser. We find that the orbiting frequency is nearly proportional to the laser power (Fig.\,\ref{fig:4}). This behavior is in sharp contrast to the property of the frequency of the laser confinement, which is proportional to the square root of the laser power~\cite{grimm2000optical,chang2010cavity}. The observed linearity is a direct proof that the orbital motion is qualitatively different from known light-induced oscillations, such as phonon lasing and the parametric instability, that are driven by radiation pressure: These phenomena show a threshold behavior with respect to the light power and a saturation at high power~\cite{vahala2009phonon,braginsky2001parametric}, neither of which is observed in the present study. Our argument is further strengthened by the fact that our experimental configuration has no mechanism of yielding rotational radiation pressure~\cite{fnote5}. We also point out that the heat generated at nanoparticles via light absorption is not likely to be the origin of the orbital motion, because its energy scale is orders of magnitude smaller than that of the orbital motion~\cite{fnote5}. 

The observed orbital motion can be regarded as a particular realization of the self-organization or the collective pattern formation of randomly moving objects that have been observed in biological, physical, and social systems~\cite{romanczuk2012active}. Phase transitions between an ensemble of randomly moving objects and their collective ordering have attracted great interests from the viewpoint of nonequilibrium statistical physics~\cite{vicsek1995novel}. As compared to other systems, which involve many objects, the present case is distinct in that a phase transition is observed only with two particles. The vortex pattern formation is of particular relevance to the present study and a theoretical model based on a two-dimensional limit cycle has been proposed~\cite{erdmann2002excitation,erdmann2005attractors}. The common feature found in these studies is that the spatial order is broken with the addition of noise, in contrast to the present study where the source of dissipation, air, is inevitable for the orbital motion to occur. Thus, we find that, although a similar theoretical model may be applicable to the present case, a qualitatively new feature has to be introduced to the model for describing the orbital motion.

To conclude, we have discovered a spontaneous orbital motion of two laser-trapped nanoparticles and demonstrated a phase transition between two Brownian particles and a pair of co-orbiting particles. The orbital motion requires the presence of air and its frequency is proportional to the laser power. Elucidating the driving mechanism of the orbital motion will be an important future task. Our work opens up a unique possibility of manipulating nanoscale objects with their cooperative dynamics. With the high controllability and the precision for observing nanoparticles, our system will be ideal for exploring the collective behaviors of few-particle systems at far-from-equilibrium. Furthermore, owing to the high sensitivity to the mass variation of nanoparticles, our system provides a new approach for nanoscale gas sensing~\cite{wang2004functional}.

\section{Supplementary Information}
\subsection{Experimental setup}
A single-frequency infrared laser at a wavelength of $\unit[1550]{nm}$ with a power of about $\unit[200]{mW}$ is focused to a beam waist of about $\unit[1.5]{\mu m}$ and is retro-reflected to form a standing-wave optical trap (an optical lattice). A piezo module attached to the retro-reflecting mirror allows us to precisely control the position of the trapped nanoparticles in the $z$ direction. The light is linearly polarized along the $x$ direction. The intensity of the trapping beam is controlled by an acousto-optic modulator (AOM). A part of the trapping beam is extracted at a polarization beam splitter and is used for detecting the motion of nanoparticles via two photodetectors. A balanced photodetector (PD1) subtracting the signal without nanoparticles from the signal with nanoparticles provides a low noise signal. A quadrant photodetector (QPD) provides the time evolution of the spatial distribution of the trapping beam. The DC voltage of the QPD is about $\unit[6]{V}$ for all the four channels. The trapped nanoparticles are imaged through an objective lens with a numerical aperture of 0.42 on a CMOS camera. For imaging, an ultraviolet (UV) light at $\unit[372]{nm}$ is overlapped with the trapping laser with a dichroic mirror and shone on nanoparticles. The infrared light scattered perpendicularly to the trapping beam by nanoparticles is monitored via an independent photodetector (PD2) that allows us to estimate the size of nanoparticles. The setup around the trap region is installed in a vacuum chamber.

\subsection{Loading nanoparticles}
We load nanoparticles into the trap in the following manner~\cite{gieseler2012subkelvin,vovrosh2017parametric,yoneda2017thermal,millen2014nanoscale}. For each implementation, we introduce a mist of ethanol including ${\rm Cu}_2{\rm O}$ nanoparticles into the vacuum chamber. When an ethanol droplet crosses a tightly focused laser beam at around the focus of the laser beam, ethanol is evaporated and nanoparticles are left trapped. 

The frequencies of the harmonic confinement of ${\rm Cu}_2{\rm O}$ nanoparticles are measured to be about $\unit[60]{kHz}$, $\unit[50]{kHz}$, and $\unit[150]{kHz}$ in the $x$, $y$, and $z$ directions, respectively, at around $\unit[400]{Pa}$, slightly above the pressure where they disappear. We infer that the disappearance indicates evaporation due to heating with laser absorption (see below).

\subsection{Estimation of the nanoparticle size}
For silica nanoparticles, previous work has demonstrated a reliable method for estimating their size from the spectral width at around a few $\unit[100]{Pa}$, which is determined by collisions with background gases~\cite{gieseler2012subkelvin,vovrosh2017parametric}. However, for ${\rm Cu}_2{\rm O}$ nanoparticles, the same method does not provide a reliable estimation of their size, mainly because they are heated via light absorption and finally disappear at around $\unit[400]{Pa}$. Therefore, we estimate the size of ${\rm Cu}_2{\rm O}$ nanoparticles in the following manner. First, we measure the size of a silica nanoparticle according to the previous method~\cite{gieseler2012subkelvin,vovrosh2017parametric} and simultaneously record the amplitude of the infrared scattering from it. Second, we record the amplitude of the infrared scattering from a ${\rm Cu}_2{\rm O}$ nanoparticle. Third, taking into account the dependence of the photon scattering cross section of a nanoparticle $\sigma$ on the refractive index $n$, we relate the amplitude of the infrared scattering with the nanoparticle size. $\sigma$ is given by the following expression:
\begin{eqnarray}
\sigma = \frac{8\pi^3|\alpha|^2}{3\epsilon_0^2\lambda^4}
\end{eqnarray}
where $\epsilon_0$ is the permittivity, $\lambda$ is the wavelength of the trapping laser, and $\alpha = 4\pi\epsilon_0 a^3(n^2-1)/(n^2+2)$ is the polarizability~\cite{cox2002experiment,jain2016direct}. The value of the refractive index of silica used for this calculation is 1.45. The mean radius of the ${\rm Cu}_2{\rm O}$ nanoparticles derived from about 60 measurements is $\unit[80]{nm}$.

\subsection{Spontaneous oscillation}
When only one particle is trapped in an optical lattice in room-temperature air, its PSD is wide-spread due to its Brownian motion. We confirm, on a CMOS camera, that in most cases the oscillation starts at the moment when nanoparticles are newly trapped. We confirm that the UV light for imaging does not have any influence on the oscillation. In the present study, we measure the orbiting frequency at the focus of the trapping beam by adjusting the position of nanoparticles with a piezo module attached to the retro-reflecting mirror.

The probability of observing the oscillation with ${\rm Cu}_2{\rm O}$ nanoparticles during many implementations is about $\unit[30]{\%}$, which we interpret as a probability of trapping more than one particle in a single site of the optical lattice. The oscillation is also observed with nanoparticles made of other materials, such as silica and ${\rm TiO}_2$, with radii of about $\unit[160]{nm}$. However, with these materials, the probabilities of observing the oscillation during many implementations are lower than with ${\rm Cu}_2{\rm O}$. We still have not understood the origin of such a difference. 

When the experiment is carried out with a single-beam optical trap, the oscillation hardly occurs and, even if it occurs, does not survive for more than a few seconds. 

\subsection{Analysis of the signal from the QPD}
Each channel of the QPD detects the interference between the trapping beam and the scattered light from nanoparticles (homodyne detection)~\cite{gieseler2012subkelvin,vovrosh2017parametric,yoneda2017thermal,millen2014nanoscale}. The length scale of the motion of nanoparticles in a trap is estimated to be about $\unit[100]{nm}$ and is much smaller than the beam diameter (about $\unit[3]{\mu m}$). Therefore, the length scale of the intensity modulation at the QPD introduced by the motion of nanoparticles should be much smaller than the size of the beam at the QPD. In such a situation, when a nanoparticle moves from the center of the beam, which we define to be the origin, to a coordinate $(x,y)$, the intensity variations at the $i$-th channel of the QPD $V_i$ are given by
\begin{eqnarray}
V_1&=-b(x+y) \\
V_2&=-b(x-y) \\
V_3&=b(x+y) \\
V_4&=b(x-y)
\end{eqnarray}
where $b$ is a numerical factor. Here we used the fact that the beam is nearly round in our experiment. Thus, we can recover the projection of the trajectory on the $xy$ plane by using the detector signals as follows:
\begin{eqnarray}
x = \frac{V_3+V_4}{2b} \\
y = \frac{V_2+V_3}{2b}
\end{eqnarray}
which are used to derive the trajectory shown in Fig.\,\ref{fig:1}(c).

In this argument, we focused on the signal from a single nanoparticle. If two nanoparticles exist, we should observe the sum of the signals from both of them. Due to a large variation in the trapped nanoparticle size as mentioned before, we infer that the observed oscillation signal is dominated by the signal from the larger nanoparticle.

\subsection{Fitting the PSD of the orbital motion}

\begin{figure}[t]
\includegraphics[width=0.95\columnwidth] {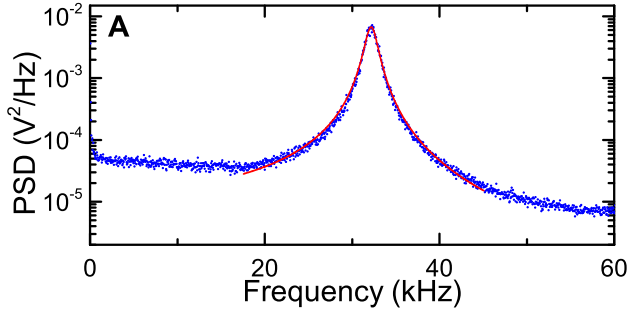}
\caption{(color online). PSD of the orbital motion at $\unit[6\times10^4]{Pa}$ and a fit on it. The Lorentzian function, shown in a solid line, fits well to the observed spectral profile. We extract both orbiting frequency and spectral width from the fit.}
\label{fig:s4}
\end{figure}

Empirically we find that the PSD of the orbital motion is well fitted by the Lorentzian function of the form
\begin{eqnarray}
S(\omega)=\frac{A}{(B^2-\omega^2)+\omega^2C^2}
\end{eqnarray}
that has been used for describing the PSD of nanoparticles at various pressures~\cite{gieseler2012subkelvin,vovrosh2017parametric}. Here, $\omega$ denotes the angular frequency and $A$, $B$, and $C$ are fitting parameters representing the magnitude, the orbiting frequency, and the spectral width, respectively. A PSD of the orbital motion at $\unit[6\times10^4]{Pa}$ and a fit to it are shown in Fig.\,\ref{fig:s4}. The quality factor of the oscillation extracted from the PSD [Fig.\,\ref{fig:1}(d)] exceeds 200, enabling us to determine the orbiting frequency with a precision of the order of $\unit[0.01]{\%}$.

\subsection{Absorption heating }

\begin{figure}[t]
\includegraphics[width=0.95\columnwidth] {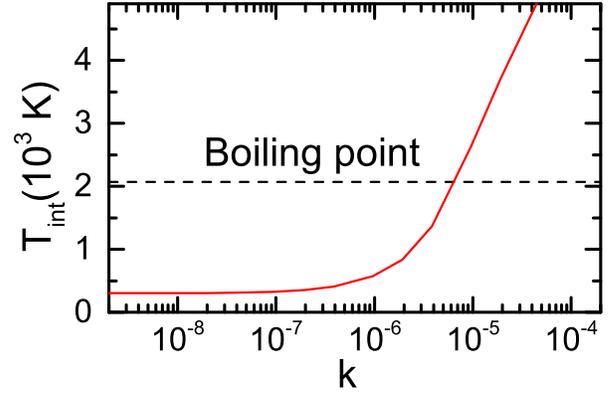}
\caption{(color online). Calculated internal temperature with respect to the extinction coefficient. The calculated internal temperature of trapped ${\rm Cu}_2{\rm O}$ nanoparticles with a radius of $\unit[80]{nm}$ for various $k$ values are plotted. The boiling point of ${\rm Cu}_2{\rm O}$ is shown by a dashed line. From this plot, we estimate the $k$ value at $\unit[1550]{nm}$ to be $7\times10^{-6}$.   }
\label{fig:s5}
\end{figure}

 The light absorption by nanoparticles can raise their internal temperature. It has been known that the heating effect is significant at low pressures~\cite{millen2014nanoscale}. However, in our working condition, we estimate that the temperature rise is minor, because nanoparticles are well cooled by surrounding air. Below we provide detailed arguments on our estimation.

\begin{figure}[t]
\includegraphics[width=0.95\columnwidth] {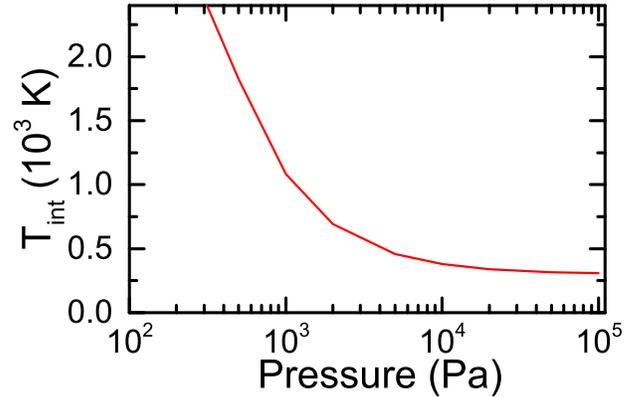}
\caption{(color online). Calculated internal temperature with respect to the pressure. The calculated internal temperature of trapped ${\rm Cu}_2{\rm O}$ nanoparticles with a radius of $\unit[80]{nm}$ for various pressures are plotted. At atmospheric pressure, the temperature increase is of the order of $\unit[1]{K}$. }
\label{fig:s6}
\end{figure}

We calculate the internal temperature of trapped ${\rm Cu}_2{\rm O}$ nanoparticles $T_{\rm int}$ according to the formalism developed in Ref.\,\cite{chang2010cavity}. For the calculation, four parameters specific to the material are needed: the complex dielectric constants at the wavelength of the trapping laser ($\unit[1550]{nm}$) and at the wavelength of blackbody radiation. The real and imaginary part of the dielectric constant are given by $n^2-k^2$  and $2nk$ , where $n$ and $k$ denote the refractive index and the extinction coefficient, respectively. We found literature values of $n$ and $k$ at around the wavelength of blackbody radiation (several $\mu$m) to be 2.3 and 0.04, respectively~\cite{querry1985optical}. The value of $n$ at around the near infrared wavelength was also found to be 2.6~\cite{karlsson1982optical}. However, we are not able to find a literature value for $k$ at around $\unit[1550]{nm}$. For estimating the value of $k$ at $\unit[1550]{nm}$, we use the fact that the trapped ${\rm Cu}_2{\rm O}$ nanoparticles disappear typically at around $\unit[400]{Pa}$ in our experiments. We interpret this disappearance as evaporation due to the temperature increase by laser absorption. Fig.\,\ref{fig:s5} shows the calculated $T_{\rm int}$ with respect to various $k$ values. The temperature of air is set to $\unit[300]{K}$. Because ${\rm Cu}_2{\rm O}$ evaporates at around $\unit[2070]{K}$, we estimate the $k$ value to be $7\times10^{-6}$. Using this value, we estimate $T_{\rm int}$ at various pressures (Fig.\,\ref{fig:s6}). We find that, at atmospheric pressure, the temperature increase is of the order of $\unit[1]{K}$, which, in principle, produces a thermal gradient around the laser-trapped nanoparticles. However, it is unlikely that such a thermal gradient can be the origin of the orbital motion with a temperature of the order of $\unit[1000]{K}$. Moreover, if we assume that the absorption heating drives the orbital motion, we find a contradiction: at low pressure, the thermal gradient increases, whereas the orbital motion is attenuated.

\begin{acknowledgments}
We thank M.\,Ueda, M.\,Kozuma, and K.\,Takeuchi for fruitful discussions. This work is supported by The Murata Science Foundation, The Mitsubishi Foundation, the Challenging Research Award and the 'Planting Seeds for Research' program from Tokyo Institute of Technology, Research Foundation for Opto-Science and Technology, JSPS KAKENHI (Grant Number 16K13857 and 16H06016), and JST PRESTO (Grant Number JPMJPR1661).
\end{acknowledgments}

\bibliographystyle{apsrev}


\begin{thebibliography}{38}
\bibitem[{fno({\natexlab{b}})}]{fnote4}
\bibinfo{note}{The oscillation is also observed with nanoparticles made of
  other materials, such as silica and TiO$_2$, with radii of about
  $\unit[160]{nm}$.}
\bibitem[{fno({\natexlab{c}})}]{fnote3}
\bibinfo{note}{Our measurement with the QPD cannot detect the motion in the $z$
  direction. Hence, even if the signals show an orbital motion in the $xy$
  plane, it might be tilted with respect to the $xy$ plane. During many
  implementations, we sometimes observe signals showing that the recovered
  trajectory is nearly a line along the $y$ direction. We interpret such a
  signal as indicating the orbital motion in the $yz$ plane.}
\bibitem[{fno({\natexlab{d}})}]{fnote1}
\bibinfo{note}{We confirm that the orbital motion is not influenced by a static
  electric field of up to 200 V/cm nor by a static magnetic field of up to 1000
  G, suggesting that nanoparticles are not ionized and the Coulomb interaction
  between them does not play a role.}
\end{thebibliography}

\end{document}